\documentclass[twocolumn,showpacs,prl,amsmath,amssymb]{revtex4}

\usepackage{graphicx}
\usepackage{dcolumn}
\usepackage{bm}

\newcommand{\gma}{$\rm{Ga_{1-x}Mn_{x}As}$}

\newcommand{\rxx}{$\rho_{\rm{xx}}$}

\newcommand{\tc}{$T_{\rm{C}}$}
\newcommand{\mni}{$\rm{Mn}_{\rm{I}}$}

\begin{document}


\title{Random telegraph noise from magnetic nanoclusters in the ferromagnetic semiconductor (Ga,Mn)As}

\author{M. Zhu}
\author{X. Li}
\author{G. Xiang}
\author{N. Samarth}
\email{nsamarth@psu.edu}
\affiliation{Department of Physics, The Pennsylvania State University, University Park PA 16802}

\date{\today}

\begin{abstract}
Measurements of the low frequency electrical noise in the ferromagnetic semiconductor (Ga,Mn)As reveal an enhanced integrated noise at low temperature. For moderate localization, we find a $1/f$ normalized power spectrum density over the entire range of temperatures studied ($4.2 \rm{K} < T < 70 \rm{K}$). However, for stronger localization and a high density of Mn interstitials, we observe Lorentzian noise spectra accompanied by random telegraph noise. Magnetic field dependence and annealing studies suggest that interstitial Mn defects couple with substitutional Mn atoms to form nanoscale magnetic clusters characterized by a net moment of $\sim 20 \mu_{\rm{B}}$ whose fluctuations modulate hole transport.

\end{abstract}

\maketitle
The canonical ferromagnetic semiconductor \gma~ is an interesting material at the confluence of several contemporary issues in condensed matter physics, ranging from spintronics to localization to correlated electron physics.\cite{Macdonald:2005rx} Point defects -- specifically Mn interstitials (\mni) and As antisites -- play an important role in the complex interplay between ferromagnetism and electronic structure in this material. Both these defects are donors that compensate the hole-mediated ferromagnetic exchange and hence suppress ferromagnetic order.\cite{ohno98} Theoretical calculations have also predicted that \mni~-- in addition to compensating holes -- may quench ferromagnetism because of antiferromagnetic coupling with neighboring substitutional Mn atoms.\cite{Blinowski:2003lr} However, there is no explicit experimental evidence for such a coupling or for any magnetic activity of \mni. In this paper, we present electrical noise measurements that suggest Mn interstitials are magnetically active in nanoscale clusters, providing a fluctuating moment that interacts with carrier transport. 

Electrical noise provides a powerful probe into the physics underlying the coupling between spin and charge transport in complex magnetic materials such as the manganites \cite{Raquet:2000:PRL,Merithew:2000:PRL,Reutler:2000:PRB,Podzorov:2000qy} and spin glasses.\cite{Weissman:1990:JAP} Our measurements of electrical noise provide new glimpses into the interplay between spin transport, hole localization and Mn interstitials in the ferromagnetic semiconductor \gma~under conditions of modest localization. We find that the resistance noise is characterized by a power spectral density (PSD) with either $1/f$ frequency dependence or a Lorentzian frequency dependence, depending on the nature of the disorder in the sample. In strong contrast with the manganites where a distinct enhancement of noise has been reported across the Curie temperature (\tc), we do not observe any measurable change in the temperature-dependent noise characteristics near \tc. Instead, we find a pronounced enhancement of the noise at low temperatures ($T \lesssim 25$ K), coinciding with increasing localization. Samples with a high density of \mni~defects show random telegraph noise (RTN) created by a small number of two-level fluctuators with an energy gap that is linear in magnetic field: we interpret this as evidence for nanoscale magnetic clusters whose spin flip fluctuations modulate transport in the Mn impurity band\cite{Burch:2006lr} and result in two possible states with different conductivity for hopping holes. In contrast to superficially similar observations in (Ga,Er)As samples containing ErAs nanoclusters,\cite{COPPINGER:1995fk} high resolution transmission electron microscopy measurements of these \gma~samples do not show any observable precipitates (such as MnAs clusters). Further, the RTN is quenched upon thermal annealing, which is known to drive \mni~to the sample surface.\cite{Macdonald:2005rx} These observations suggest that \mni~are actively involved in a fluctuating moment that consists of only a few Mn atoms.   

\gma~epilayers are fabricated using low temperature molecular beam epitaxy on (001) semi-insulating, epiready GaAs substrates following the growth of a 170 nm thick high temperature buffer of GaAs epilayer. We selected three samples for noise measurements, with the following (as grown) characteristics determined using secondary ion mass spectrometry, room temperature Hall effect (up to 14 T) and superconducting quantum interference device magnetometry: sample A has a thickness of 30 nm, $x \sim 0.06$, a room temperature hole density $p \sim 5.6 \times 10^{19} \rm{cm}^{-3}$ and \tc$= 55$K;  sample B has a thickness of 30 nm, $x \sim 0.066$,$p \sim 4.4 \times 10^{19} \rm{cm}^{-3}$ at 300 K, and \tc$= 50$K ; sample C has a thickness of 50 nm, $x \sim 0.08$, $p \sim 1 \times 10^{21} \rm{cm}^{-3}$ at 300 K and \tc$= 100$K. These data indicate that samples A and B have similar net Mn concentration, but that the latter has a larger \mni~density. The higher level of disorder in sample B results in significantly more localization at low temperatures: the resistivity of samples A and B at 4.2 K is 12 m$\Omega$.cm and 60 m$\Omega$.cm, respectively. We also measured a control sample consisting of 100 nm thick epilayer of  p-GaAs (doped with Be) with  $p \sim 10^{21} \rm{cm}^{-3}$ at 300 K.  Samples are patterned using a wet mesa etch into a 50 $\mu$m wide Hall-bar-like geometry along the [110] crystalline direction, with a spacing of 200 $\mu$m between adjacent longitudinal voltage probes (contacted with In). 

Resistance noise measurements are carried out using a five-probe ac scheme,\cite{Scofield:1987:Rev_Sci_Ins} with a lock-in amplifier (SR830) providing the excitation current and demodulating the fluctuation signal, with the output fed into a spectrum analyzer (SR785). The resistances of the balancing resistors are at least 10 x the sample resistance to minimize their contribution to the overall noise. In typical ac noise measurements, the excitation frequency is chosen to lie in the eye of the noise figure of the preamplifier (100Hz-1KHz). However, in \gma, we select an excitation frequency $f_{0} = 43$Hz because of a strong frequency dependence of resistance at high frequencies. The resistance fluctuations modulate the carriers to produce noise sidebands, which are demodulated by phase-sensitive detection on the lock-in amplifier. Thus, the validity of ac measurements is limited to $f < f_{0}$/2, where $f$ is the frequency of the noise spectrum \cite{Scofield:1987:Rev_Sci_Ins}. The output low-pass filter of the lock-in is set to have a 3 ms time constant with a rolloff of 24dB/Oct \cite{Bid:2003:PRB}. Under these conditions, the intrinsic noise from the instrument shows a flat spectrum up to 12Hz, setting the upper limit of our spectra. The PSD of the background noise from the instrument is $2.7\times10^{-16}$V$^{2}$/Hz at room temperature. A quadratic dependence of PSD on the applied bias voltage indicates that the measured voltage fluctuations indeed stem from the fluctuations of sample resistance \cite{Raychaudhuri:2002:Cur_Opin}. The output voltage from the lock-in is then set in the range 3V - 5V rms which gives a current of 5$\mu$A-10$\mu$A depending on balancing resistors and sample resistance. The selected current density ($\sim$10$^{2}$ A/cm$^{2}$) is high enough to distinguish the excess noise from background noise, but low enough to avoid substantial heating.

\begin{figure}[t]
\begin{center}
\includegraphics{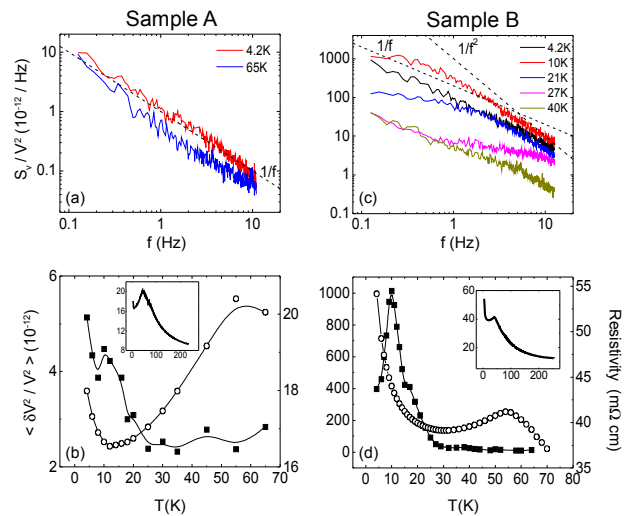} 
\caption{(color online). (a) Normalized PSD of sample A at two temperatures showing $1/f$ dependence (dashed line). (b) Temperature dependence of the integrated normalized PSD (left) and \rxx~ (right) for sample A. Inset shows the \rxx$(T)$ over an expanded temperature range. (c) Normalized PSD of sample B at different temperatures. Dashed line shows $1/f$ and $1/f^{2}$ relations. Strong deviations of PSD from $1/f$ are clear over range 6 K $\lesssim T \lesssim$ 29 K. (d) Temperature dependence of the integrated normalized PSD (left) and  \rxx~(right) for sample B. Inset shows the \rxx$(T)$ over an expanded temperature range.}
\end{center}
\end{figure}

Noise measurements in the higher conductivity control sample and the high \tc~ sample (C) do not yield significant noise, perhaps as a limitation of our current measurement set up. Hence, we focus our discussion on the lower conductivity samples A and B where the noise signal is readily measured. Figure 1 (a) shows the normalized PSD of the voltage fluctuations ($S_{V}/V^{2}$) for sample A at different temperatures, demonstrating $1/f$-resistance noise over a wide temperature range.  We recall that a single fluctuator with a relaxation time $\tau$, yields a Lorentzian PSD with the form $S(f)\propto 2\tau/1+(2\pi f\tau)^{2}$. An ensemble of fluctuators with a broad distribution of relaxation times then produces a $1/f$ spectrum due to the superposition of many Lorentzians.\cite{Raychaudhuri:2002:Cur_Opin} The magnitude of $1/f$ noise is characterized using the phenomenological equation\cite{Dutta:1981:Rev_Mod_Phy}:
\begin{equation} 
\frac{S_{V}(f)}{V^{2}}=\frac{\gamma}{n_{c}\Omega f^{\alpha}},
\end{equation}
where $\gamma$ is the Hooge parameter,  $V$ is the voltage across the sample, $n_{c}$ is the carrier density and $\Omega$ is the sample volume. Using $n_{c} \sim 1\times10^{20}$/cm$^{3}$, we obtain $\gamma = 3\times10^{-2}$, comparable to typical values reported in metals ($\sim 10^{-2}$-$10^{-3}$) \cite{Kogan}. The exponent $\alpha$ varies from 0.96 to 1.18 over the temperature range $4.2 \rm {K} < T < 65$K. 

The temperature-dependence of the relative rms fluctuation $\langle\delta V^{2}/V^{2}\rangle$ is obtained by integrating the normalized noise $S_{V}/V^{2}$ over the range 125 mHz -- 11 Hz. This is shown in fig.1 (b), along with the temperature dependence of the longitudinal resistivity (\rxx). Our measurements do not indicate any detectable correlation between the noise amplitude and \tc (where \rxx~has a peak). This contrasts with the manganites where the integrated noise has a peak at \tc, attributed to a percolative phase transition between the charge-ordered and FM phases.\cite{Podzorov:2000qy}. We speculate that in \gma, contributions to resistance fluctuations from possible mixed phases may be overshadowed by the effect of disorder. At low temperatures where \rxx~goes through a minimum, we observe an increase in noise amplitude presumably from the increased localization of holes. 

We now address the noise characteristics of sample B, which has a \tc~and net Mn concentration similar to that of sample A, but has a higher \mni~density. As in sample A, there is no evidence for a correlation between noise characteristics and \tc, and there is a marked enhancement of noise at low temperatures. In contrast with sample A, however, we find clear deviations from $1/f$-noise over the temperature range from 6 K to 29 K (Fig.1(c)).  Further, the enhancement of noise at low temperatures is far more pronounced compared with sample A: as shown in Fig. 1 (d), the relative rms fluctuation peaks at 10 K, reaching a value that is almost two orders of magnitude higher than that above 30 K. At the lowest temperature measured (4.2 K), $1/f$ behavior is restored, accompanied by a clear drop in the noise amplitude. We attribute the deviation of PSD from $1/f$ to the dominance of a few fluctuators, each of which has a different relaxation time.  The $1/f^{2}$ tail observed at higher frequencies in Fig.1(c) is evidence of such Lorentzian spectra and is corroborated by time-domain measurements (Fig.2(a)) that show RTN, with $\delta V/V \sim 0.01\%$.

\begin{figure}[t]
\begin{center}
\includegraphics{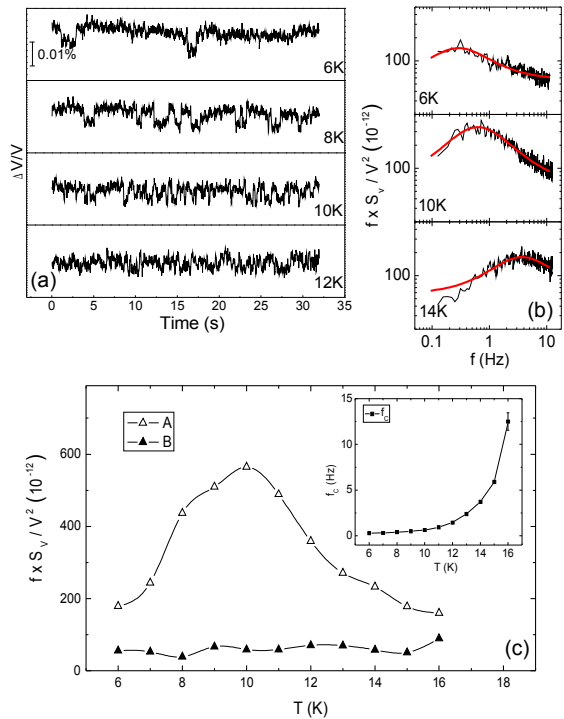} 
\caption{(color online). (a) Time-dependent voltage fluctuations in sample B, showing RTN. (b) $fS_{V}(f)/V^{2}$ at different temperatures in sample B. Solid lines are fits to Lorentzian spectra with a superposition of a $1/f$ background (Eq. 2). (c) Relative strengths of the Lorentzian term (A) and the $1/f$ term (B) obtained by fits to Eq. 2. Inset shows the corner frequency $f_{c}$ vs. $T$ } 
\end{center}
\end{figure}

We model the observed RTN as a simple two-level fluctuator (TLF) in which the barrier heights to escape from one state to another are $E+\delta E$ and $E-\delta E$ respectively. The average time spent in each state is given by $\tau_{i}= (2\pi f_{0,i})^{-1}$exp$(E\pm\delta E/k_{B}T)$. The overall relaxation time of the whole system is $1/\tau_{c}=1/\tau_{1}+1/\tau_{2}$. The normalized PSD is then given by a Lorentzian term plus a small $1/f$ background:
\begin{equation} 
\frac{S_{V}(f)}{V^{2}}=\frac{2\pi\tau_{c}A}{1+(2\pi\tau_{c})^{2}f^{2}}+ \frac{B}{f}=\frac{f_{c}A}{f_{c}^{2}+f^{2}}+ \frac{B}{f},
\end{equation}
where A and B are the relative strengths of the $1/f$ and $1/f^2$ contributions, respectively, and $f_{c}=1/(2\pi\tau_{c})$ is the corner frequency of the Lorentzian. These parameters can be extracted by fitting the frequency dependence of $fS_{V}/V^{2}$ \cite{Muller:2006:PRL} (Fig. 2 (b)). 

Figure 2 (c) shows the temperature dependence of the fitting parameters A, B and $f_{c}$, demonstrating that the temperature variation of the noise spectrum in Fig. 1 (d) is dominated by that of the Lorentzian contribution. The corner frequency $f_{c}$ shifts monotonically toward higher frequencies with increasing temperature, indicating phonon activated processes. Further determination of the thermal attempt frequency $f_{0}$ and barrier heights is found to be difficult due to the limited temperature range and exponential dependence of $f_{0}$ on $1/T$. 

To further explore the origin of the RTN, we measure the magnetic field-dependence of the noise spectra with an in-plane magnetic field $\vec{H} || [110]$ (Fig. 3(a)). These measurements are compared with the switching of the magnetization ($\vec{M}$), tracked using the giant planar Hall effect (GPHE).\cite{Tang:2003fj}) The biaxial anisotropy in \gma~results in easy axes close to the [100] and [010] directions; at $H = 0$, $\vec{M} || [\bar{1}00]$, corresponding to one of the four energy minima. The enhancement and subsequent sudden quenching of noise at low field ($0 < H \lesssim 90$Oe) signals the nucleation of small domains, followed by their coalescence into a single macroscopic domain with $\vec{M} || [010]$. The second switching event -- when $\vec{M}$ rotates from [010] to [100] -- is also accompanied by the enhancement and subsequent suppression of noise, although the changes are less abrupt because the magnetization reversal process is dominated by coherent rotation rather than domain nucleation. At higher fields ($0.5 \rm{T} < H < 2 \rm{T}$), $\vec{M}$ gradually rotates coherently towards $\vec{H}$, accompanied by a striking enhancement in integrated noise and RTN, before the PSD reverts to $1/f$-behavior at $H > 3$T.  A fit to eq. 2 shows that the Lorentzian contribution peaks at $H = 1.2$T. Further, in time domain, the dwelling times of RTN for up and down states change systematically with field (Fig. 3(b)), implying field-dependent energy barriers. The field dependence of the noise amplitude and energy barriers both suggest a magnetic origin of the TLF, rather than domain fluctuations. 

\begin{figure}[t]
\begin{center}
\includegraphics{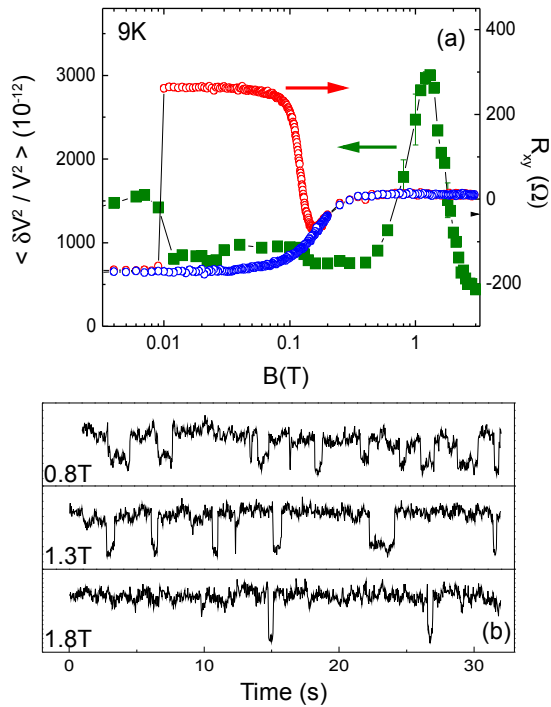} 
\caption{(color online). (a) Magnetic field-dependence of the relative rms fluctuation (solid squares) at $T = 9$K. For comparison, we also plot the GPHE with red (blue) circles for up (down) field sweeps. (b) Time traces of RTN at different fields, also at $T = 9$K.}
\end{center}
\end{figure}

A possible origin for the RTN is spin-dependent scattering from nanoscale magnetic clusters that fluctuate between two possible orientations determined either by an external magnetic field $H$ or magnetocrystalline anisotropy.\cite{COPPINGER:1995fk} 
In this case, the barrier heights may be written in terms of a field-independent and a field-dependent contribution. The average time in each states is: $\tau_{i}=(2\pi f_{0,i})^{-1}$exp$(E\pm\delta E + E_{i}(H)/k_{B}T)$. The ratio of dwelling time is then: $\tau_{1}/\tau_{2}$=exp$(2\delta E+\Delta E(H)/k_{B}T)$, where $\tau_{1}$ and $\tau_{2}$ is the average time spent in the 'spin up' and 'spin down' orientations, respectively. For a cluster of magnetic moment $m$, we expect $\Delta E(H) = mH$. By measuring the mean dwelling time in each of the two states (averaged over 10 minutes) at different magnetic fields and plotting $k_{B}T\ln\tau_{1}/\tau_{2}$ as a function of magnetic field, we obtain a linear field dependence, from which $ m \sim 20\mu_{B}$ is extracted. This suggests evidence for nanoscale magnetic clusters with a few Mn atoms. Since the RTN decreases significantly upon annealing (Fig. 4(b)), we infer that Mn interstitials must play a direct role in these clusters and speculate that -- as suggested by theory \cite{Blinowski:2003lr}  -- Mn interstitials couple antiferromagnetically with substitutional Mn atoms to form nanoscale clusters with uncompensated spins. These clusters may coherently influence the hopping conduction of holes occupying the Mn impurity band, possibly achieving two states with different conductivities. The onset of magnetic correlation in such clusters at low temperature would explain the pronounced enhancement of noise at $T \sim 10$K. At even lower temperatures ($T \lesssim 10$ K), the fluctuations of these clusters are frozen out and carrier hopping conduction is not affected. Equivalently, at higher temperatures, spin flip fluctuations of these clusters are frozen out by the application of an external magnetic field, quenching the RTN as a result of the increased barrier heights.

\begin{figure}[t]
\begin{center}
\includegraphics{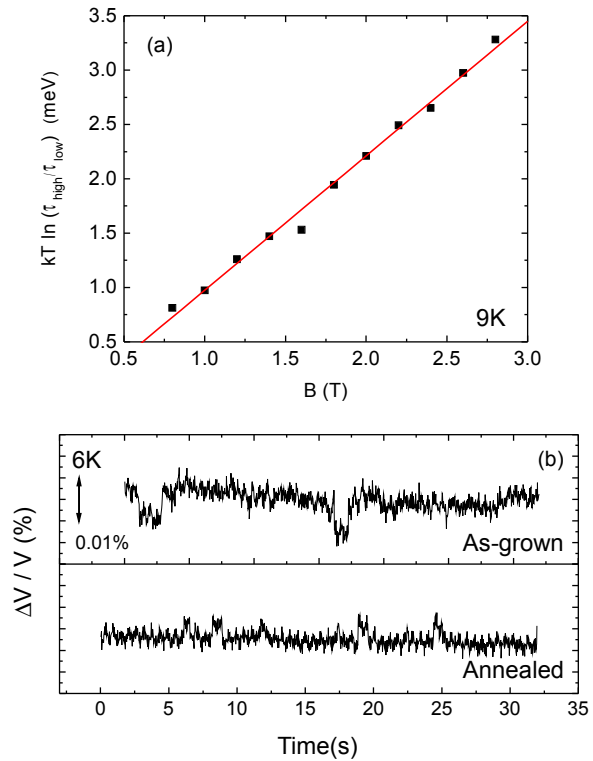} 
\caption{(color online). (a) Magnetic field dependence of energy difference $\Delta E$ between the two states of the TLF in sample B at $T = 6$K. (b) Comparison between RTN at $T = 6$K in as-grown and annealed pieces of sample B.}
\end{center}
\end{figure}

In summary, electrical noise measurements shed new light into the interplay between transport, localization and magnetism in \gma. Surprisingly, we do not observe any change in noise in the vicinity of the Curie temperature, ruling out a percolative transition with coexisting phases as in the manganites; instead, a striking enhancement of noise is observed at low temperatures with increasing carrier localization. In samples with a large density of Mn interstitial defects, we find evidence for TLFs whose behavior with magnetic field (and annealing) suggest that these defects are incorporated into magnetically-active nanoscale clusters.

We thank Stephan von Molnar, Peng Xiong and Jens M\"uller for helpful discussions and technical advice, and Ben-Li Sheu and Xianglin Ke for high field Hall measurements. Work supported by grant numbers ONR N0014-05-1-0107 and NSF DMR-0305238. We also acknowledge the use of the NSF funded NNIN facility at Penn State University.

\newpage

\begin{thebibliography}{17}
\expandafter\ifx\csname natexlab\endcsname\relax\def\natexlab#1{#1}\fi
\expandafter\ifx\csname bibnamefont\endcsname\relax
  \def\bibnamefont#1{#1}\fi
\expandafter\ifx\csname bibfnamefont\endcsname\relax
  \def\bibfnamefont#1{#1}\fi
\expandafter\ifx\csname citenamefont\endcsname\relax
  \def\citenamefont#1{#1}\fi
\expandafter\ifx\csname url\endcsname\relax
  \def\url#1{\texttt{#1}}\fi
\expandafter\ifx\csname urlprefix\endcsname\relax\def\urlprefix{URL }\fi
\providecommand{\bibinfo}[2]{#2}
\providecommand{\eprint}[2][]{\url{#2}}

\bibitem[{\citenamefont{Macdonald et~al.}(2005)\citenamefont{Macdonald,
  Schiffer, and Samarth}}]{Macdonald:2005rx}
\bibinfo{author}{\bibfnamefont{A.~H.} \bibnamefont{Macdonald}},
  \bibinfo{author}{\bibfnamefont{P.}~\bibnamefont{Schiffer}}, \bibnamefont{and}
  \bibinfo{author}{\bibfnamefont{N.}~\bibnamefont{Samarth}},
  \bibinfo{journal}{Nature (Materials)} \textbf{\bibinfo{volume}{4}},
  \bibinfo{pages}{195} (\bibinfo{year}{2005}).

\bibitem[{\citenamefont{Ohno}(1998)}]{ohno98}
\bibinfo{author}{\bibfnamefont{H.}~\bibnamefont{Ohno}},
  \bibinfo{journal}{Science} \textbf{\bibinfo{volume}{281}}
  (\bibinfo{year}{1998}).

\bibitem[{\citenamefont{Blinowski and Kacman}(2003)}]{Blinowski:2003lr}
\bibinfo{author}{\bibfnamefont{J.}~\bibnamefont{Blinowski}} \bibnamefont{and}
  \bibinfo{author}{\bibfnamefont{P.}~\bibnamefont{Kacman}},
  \bibinfo{journal}{Phys. Rev. B} \textbf{\bibinfo{volume}{67}},
  \bibinfo{pages}{121204} (\bibinfo{year}{2003}).

\bibitem[{\citenamefont{Raquet et~al.}(2000)\citenamefont{Raquet, Anane, Wirth,
  Xiong, and von molnar}}]{Raquet:2000:PRL}
\bibinfo{author}{\bibfnamefont{B.}~\bibnamefont{Raquet}},
  \bibinfo{author}{\bibfnamefont{A.}~\bibnamefont{Anane}},
  \bibinfo{author}{\bibfnamefont{S.}~\bibnamefont{Wirth}},
  \bibinfo{author}{\bibfnamefont{P.}~\bibnamefont{Xiong}}, \bibnamefont{and}
  \bibinfo{author}{\bibfnamefont{S.}~\bibnamefont{von Molnar}},
  \bibinfo{journal}{Phys. Rev. Lett.} \textbf{\bibinfo{volume}{84}},
  \bibinfo{pages}{4485 } (\bibinfo{year}{2000}).

\bibitem[{\citenamefont{Merithew et~al.}(2000)\citenamefont{Merithew, Weissman,
  Hess, Spradling, Nowak, O'donnell, Eckstein, Tokura, and
  Tomioka}}]{Merithew:2000:PRL}
\bibinfo{author}{\bibfnamefont{R.}~\bibnamefont{Merithew}},
  \bibinfo{author}{\bibfnamefont{M.}~\bibnamefont{Weissman}},
  \bibinfo{author}{\bibfnamefont{F.}~\bibnamefont{Hess}},
  \bibinfo{author}{\bibfnamefont{P.}~\bibnamefont{Spradling}},
  \bibinfo{author}{\bibfnamefont{E.}~\bibnamefont{Nowak}},
  \bibinfo{author}{\bibfnamefont{J.}~\bibnamefont{O'donnell}},
  \bibinfo{author}{\bibfnamefont{J.}~\bibnamefont{Eckstein}},
  \bibinfo{author}{\bibfnamefont{Y.}~\bibnamefont{Tokura}}, \bibnamefont{and}
  \bibinfo{author}{\bibfnamefont{Y.}~\bibnamefont{Tomioka}},
  \bibinfo{journal}{Phys. Rev. Lett.} \textbf{\bibinfo{volume}{84}},
  \bibinfo{pages}{3442 } (\bibinfo{year}{2000}).

\bibitem[{\citenamefont{Reutler et~al.}(2000)\citenamefont{Reutler, Bensaid,
  Herbstritt, Hofener, Marx, and Gross}}]{Reutler:2000:PRB}
\bibinfo{author}{\bibfnamefont{P.}~\bibnamefont{Reutler}},
  \bibinfo{author}{\bibfnamefont{A.}~\bibnamefont{Bensaid}},
  \bibinfo{author}{\bibfnamefont{F.}~\bibnamefont{Herbstritt}},
  \bibinfo{author}{\bibfnamefont{C.}~\bibnamefont{Hofener}},
  \bibinfo{author}{\bibfnamefont{A.}~\bibnamefont{Marx}}, \bibnamefont{and}
  \bibinfo{author}{\bibfnamefont{R.}~\bibnamefont{Gross}},
  \bibinfo{journal}{Phys. Rev. B} \textbf{\bibinfo{volume}{62}},
  \bibinfo{pages}{11619 } (\bibinfo{year}{2000}).

\bibitem[{\citenamefont{Podzorov et~al.}(2000)\citenamefont{Podzorov, Uehara,
  Gershenson, Koo, and Cheong}}]{Podzorov:2000qy}
\bibinfo{author}{\bibfnamefont{V.}~\bibnamefont{Podzorov}},
  \bibinfo{author}{\bibfnamefont{M.}~\bibnamefont{Uehara}},
  \bibinfo{author}{\bibfnamefont{M.~E.} \bibnamefont{Gershenson}},
  \bibinfo{author}{\bibfnamefont{T.~Y.} \bibnamefont{Koo}}, \bibnamefont{and}
  \bibinfo{author}{\bibfnamefont{S.~W.} \bibnamefont{Cheong}},
  \bibinfo{journal}{Phys. Rev. B} \textbf{\bibinfo{volume}{61}},
  \bibinfo{pages}{R3784} (\bibinfo{year}{2000}).

\bibitem[{\citenamefont{Weissman and Israeloff}(1990)}]{Weissman:1990:JAP}
\bibinfo{author}{\bibfnamefont{M.}~\bibnamefont{Weissman}} \bibnamefont{and}
  \bibinfo{author}{\bibfnamefont{N.}~\bibnamefont{Israeloff}},
  \bibinfo{journal}{J. Appl. Phys.} \textbf{\bibinfo{volume}{67}},
  \bibinfo{pages}{4884 } (\bibinfo{year}{1990}).

\bibitem[{\citenamefont{Burch et~al.}(2006)\citenamefont{Burch, Shrekenhamer,
  Singley, Stephens, Sheu, Kawakami, Schiffer, Samarth, Awschalom, and
  Basov}}]{Burch:2006lr}
\bibinfo{author}{\bibfnamefont{K.~S.} \bibnamefont{Burch}},
  \bibinfo{author}{\bibfnamefont{D.~B.} \bibnamefont{Shrekenhamer}},
  \bibinfo{author}{\bibfnamefont{E.~J.} \bibnamefont{Singley}},
  \bibinfo{author}{\bibfnamefont{J.}~\bibnamefont{Stephens}},
  \bibinfo{author}{\bibfnamefont{B.~L.} \bibnamefont{Sheu}},
  \bibinfo{author}{\bibfnamefont{R.~K.} \bibnamefont{Kawakami}},
  \bibinfo{author}{\bibfnamefont{P.}~\bibnamefont{Schiffer}},
  \bibinfo{author}{\bibfnamefont{N.}~\bibnamefont{Samarth}},
  \bibinfo{author}{\bibfnamefont{D.~D.} \bibnamefont{Awschalom}},
  \bibnamefont{and} \bibinfo{author}{\bibfnamefont{D.~N.} \bibnamefont{Basov}},
  \bibinfo{journal}{Phys. Rev. Lett.} \textbf{\bibinfo{volume}{97}},
  \bibinfo{pages}{087208} (\bibinfo{year}{2006}).

\bibitem[{\citenamefont{Coppinger et~al.}(1995)\citenamefont{Coppinger, Genoe,
  Maude, Gennser, Portal, Singer, Rutter, Taskin, Peaker, and
  Wright}}]{COPPINGER:1995fk}
\bibinfo{author}{\bibfnamefont{F.}~\bibnamefont{Coppinger}},
  \bibinfo{author}{\bibfnamefont{J.}~\bibnamefont{Genoe}},
  \bibinfo{author}{\bibfnamefont{D.~K.} \bibnamefont{Maude}},
  \bibinfo{author}{\bibfnamefont{U.}~\bibnamefont{Gennser}},
  \bibinfo{author}{\bibfnamefont{J.~C.} \bibnamefont{Portal}},
  \bibinfo{author}{\bibfnamefont{K.~E.} \bibnamefont{Singer}},
  \bibinfo{author}{\bibfnamefont{P.}~\bibnamefont{Rutter}},
  \bibinfo{author}{\bibfnamefont{T.}~\bibnamefont{Taskin}},
  \bibinfo{author}{\bibfnamefont{A.~R.} \bibnamefont{Peaker}},
  \bibnamefont{and} \bibinfo{author}{\bibfnamefont{A.~C.}
  \bibnamefont{Wright}}, \bibinfo{journal}{Phys. Rev. Lett.}
  \textbf{\bibinfo{volume}{75}}, \bibinfo{pages}{3513} (\bibinfo{year}{1995}).

\bibitem[{\citenamefont{Scofield}(1987)}]{Scofield:1987:Rev_Sci_Ins}
\bibinfo{author}{\bibfnamefont{J.}~\bibnamefont{Scofield}},
  \bibinfo{journal}{Rev. Sci. Instr.} \textbf{\bibinfo{volume}{58}},
  \bibinfo{pages}{985 } (\bibinfo{year}{1987}).

\bibitem[{\citenamefont{Bid et~al.}(2003)\citenamefont{Bid, Guha, and
  Raychaudhuri}}]{Bid:2003:PRB}
\bibinfo{author}{\bibfnamefont{A.}~\bibnamefont{Bid}},
  \bibinfo{author}{\bibfnamefont{A.}~\bibnamefont{Guha}}, \bibnamefont{and}
  \bibinfo{author}{\bibfnamefont{A.}~\bibnamefont{Raychaudhuri}},
  \bibinfo{journal}{Phys. Rev. B} \textbf{\bibinfo{volume}{67}},
  \bibinfo{pages}{174415} (\bibinfo{year}{2003}).

\bibitem[{\citenamefont{Raychaudhuri}(2002)}]{Raychaudhuri:2002:Cur_Opin}
\bibinfo{author}{\bibfnamefont{A.}~\bibnamefont{Raychaudhuri}},
  \bibinfo{journal}{Current Opinions in Solid State and Materials Science}
  \textbf{\bibinfo{volume}{6}}, \bibinfo{pages}{67 } (\bibinfo{year}{2002}).

\bibitem[{\citenamefont{Dutta and Horn}(1981)}]{Dutta:1981:Rev_Mod_Phy}
\bibinfo{author}{\bibfnamefont{P.}~\bibnamefont{Dutta}} \bibnamefont{and}
  \bibinfo{author}{\bibfnamefont{P.}~\bibnamefont{Horn}},
  \bibinfo{journal}{Rev. Mod. Phys.} \textbf{\bibinfo{volume}{53}},
  \bibinfo{pages}{497 } (\bibinfo{year}{1981}).

\bibitem[{\citenamefont{Sh.Kogan}(1996)}]{Kogan}
\bibinfo{author}{\bibnamefont{Sh.Kogan}}, \emph{\bibinfo{title}{Electronic
  noise and fluctuations in solids}} (\bibinfo{publisher}{Cambridge University
  Press}, \bibinfo{year}{1996}).

\bibitem[{\citenamefont{Muller et~al.}(2006)\citenamefont{Muller, von molnar,
  Ohno, and Ohno}}]{Muller:2006:PRL}
\bibinfo{author}{\bibfnamefont{J.}~\bibnamefont{M\"uller}},
  \bibinfo{author}{\bibfnamefont{S.}~\bibnamefont{von molnar}},
  \bibinfo{author}{\bibfnamefont{Y.}~\bibnamefont{Ohno}}, \bibnamefont{and}
  \bibinfo{author}{\bibfnamefont{H.}~\bibnamefont{Ohno}},
  \bibinfo{journal}{Phys. Rev. Lett.} \textbf{\bibinfo{volume}{96}},
  \bibinfo{pages}{186601} (\bibinfo{year}{2006}).

\bibitem[{\citenamefont{Tang et~al.}(2003)\citenamefont{Tang, Kawakami,
  Awschalom, and Roukes}}]{Tang:2003fj}
\bibinfo{author}{\bibfnamefont{H.~X.} \bibnamefont{Tang}},
  \bibinfo{author}{\bibfnamefont{R.~K.} \bibnamefont{Kawakami}},
  \bibinfo{author}{\bibfnamefont{D.~D.} \bibnamefont{Awschalom}},
  \bibnamefont{and} \bibinfo{author}{\bibfnamefont{M.~L.}
  \bibnamefont{Roukes}}, \bibinfo{journal}{Phys. Rev. Lett.}
  \textbf{\bibinfo{volume}{90}}, \bibinfo{pages}{107201}
  (\bibinfo{year}{2003}).

\end{thebibliography}

\end{document}